\documentclass[aps,amssymb,amsmath,pra,showpacs,superscriptaddress]{revtex4}
\usepackage{latexsym}
\usepackage{amssymb}
\usepackage{graphicx,epstopdf}

\usepackage{color}
\date{\today}
\begin{document}
\title{Quantum key distribution using intra-particle entanglement}

\author{S. Adhikari}
\email{tapisatya@gmail.com}
\affiliation{Indian Institute of Technology, Jodhpur, India}

\author{Dipankar Home}
\email{dhome@bosemain.boseinst.ac.in}
\affiliation{CAPSS, Dept. of Physics, Bose Institute, Salt Lake,
Kolkata-700091, India}

\author{A. S. Majumdar}
\email{archan@bose.res.in}
\affiliation{S. N. Bose National Centre for Basic Sciences,
Salt Lake, Kolkata 700 098, India}

\author{A. K. Pan}
\email{akp@math.cm.is.nagoya-u.ac.jp}
\affiliation{Graduate School of Information Science, Nagoya University, Chikusa-ku, Nagoya 464-8601, Japan}

\author{Akshata Shenoy H.}
\email{akshata@ece.iisc.ernet.in}
\affiliation{ECE Dept., IISc, Bangalore, India}

\author{R. Srikanth}
\email{srik@poornaprajna.org}
\affiliation{Poornaprajna Institute of Scientific Research, Bangalore, India}

\begin{abstract}
We  propose the  use  of intra-particle  entanglement  to enhance  the
security  of a practical  implementation of  the Bennett-Brassard-1984
(BB84) quantum key distribution scheme. Intra-particle entanglement is
an attractive  resource since  it can be  easily generated  using only
linear optics. Security is studied  under a simple model of incoherent
attack  for protocols  involving  two or  all  five mutually  unbiased
bases.  In terms of efficiency  of secret key generation and tolerable
error rate, the latter is found to be superior to the former.  We find
that states that allow secrecy distillation are necessarily entangled,
though  they  may  be  local.   Since more  powerful  attacks  by  Eve
obviously  exist,  our result  implies  that  security  is a  strictly
stronger condition than entanglement for these protocols.
\end{abstract}

\date{}

\maketitle

\section{Introduction}

Quantum  key  distribution  (QKD)  protocols  \cite{gisin}  allow  two
distant  parties, traditionally  called Alice  and Bob,  to  produce a
shared random bit string consisting of 0's and 1's known only to them,
which can be used as a  key to encrypt and decrypt messages.  Based on
fundamental    principles   such    as   the    no-cloning   principle
\cite{wootters}, in  quantum physics, QKD  provides an unconditionally
secure  way to distribute  random keys  through insecure  channels. In
1984, the  well known Bennett-Brassard (BB84) protocol,  the first QKD
protocol, was  proposed \cite{bennett}.  The  first entanglement based
QKD  protocol  was  proposed  by  Ekert in  1991  \cite{ekert},  where
violation of a  local realist inequality \cite{bell} is  used to check
the security of QKD.   Subsequently, there have been several important
theoretical  advancements   in  understanding  the   relation  between
security and  nonlocality \cite{scagis1,scagis2,acin} and experimental
demonstrations  of BB84  and  other QKD  protocols  were discussed  in
\cite{lucamarini, bruss, wang, kraus, lo,adachi}.  In this connection,
the  violation of  the Bell's  inequality  was first  monitored in  an
experiment by  Jennewein et. al.\cite{jennewein},  but no quantitative
measure of  security was derived  from the observed  violation. Later,
Ling  et.  al.  \cite{ling}  performed an  experiment on  entanglement
based QKD, in  which the violation of Bell-CHSH  inequality is used to
also quantify  the degree  of security according  to the  criterion of
Ref. \cite{acin}.

In  this work,  we provide  a  QKD scheme  in which  path-polarization
entanglement  in  a  single  particle (also  known  as  intra-particle
entanglement) is  exploited.  The  four single-qubit symbol  states of
standard  BB84  are  replaced  in our  proposal  with  single-particle
states.    As  we   show   later,  single-particle   path-polarization
entanglement  offers security against  some side-channel  attacks that
can render vulnerable a  practical implementation of BB84, even though
BB84 is theoretically unbreakable \cite{sp00}.

Theoretically,  we assume  that  the  stations of  Alice  and Bob  are
private spaces that are inaccessible to the outside world. However, in
practice, a side-channel attack can be launched because their stations
are not entirely isolated. Such  attacks rely on flaws in the devices,
which may potentially  have been rigged by eavesdropper  Eve to reveal
information  about settings  and even  outcomes.  Examples  of typical
side channels are timing information on the devices used, observations
of  power  consumption  or  electromagnetic leaks  bearing  some  heat
signature of  devices, or  even a click  sound produced by  an optical
element.
 
Intra-particle   entanglement   was   discussed   by   Basu   et   al.
\cite{Home-Kar-PLA-2001}  using  a  Mach-Zehnder type  interferometric
set-up for  demonstrating the violation of  non-contextuality, and the
actual experiment  using single neutrons was performed  by Hasegawa et
al.  \cite{hasegawa-nature-2003}.  Here  we propose that the protocols
for cryptography presented here can be performed practically using the
same  approach by  appropriately replacing  path-spin  entanglement of
neutrons with path-polarization entanglement of photons.

\section{Intra-particle entanglement}

\begin{figure}[!ht]
\includegraphics[width=10.0cm]{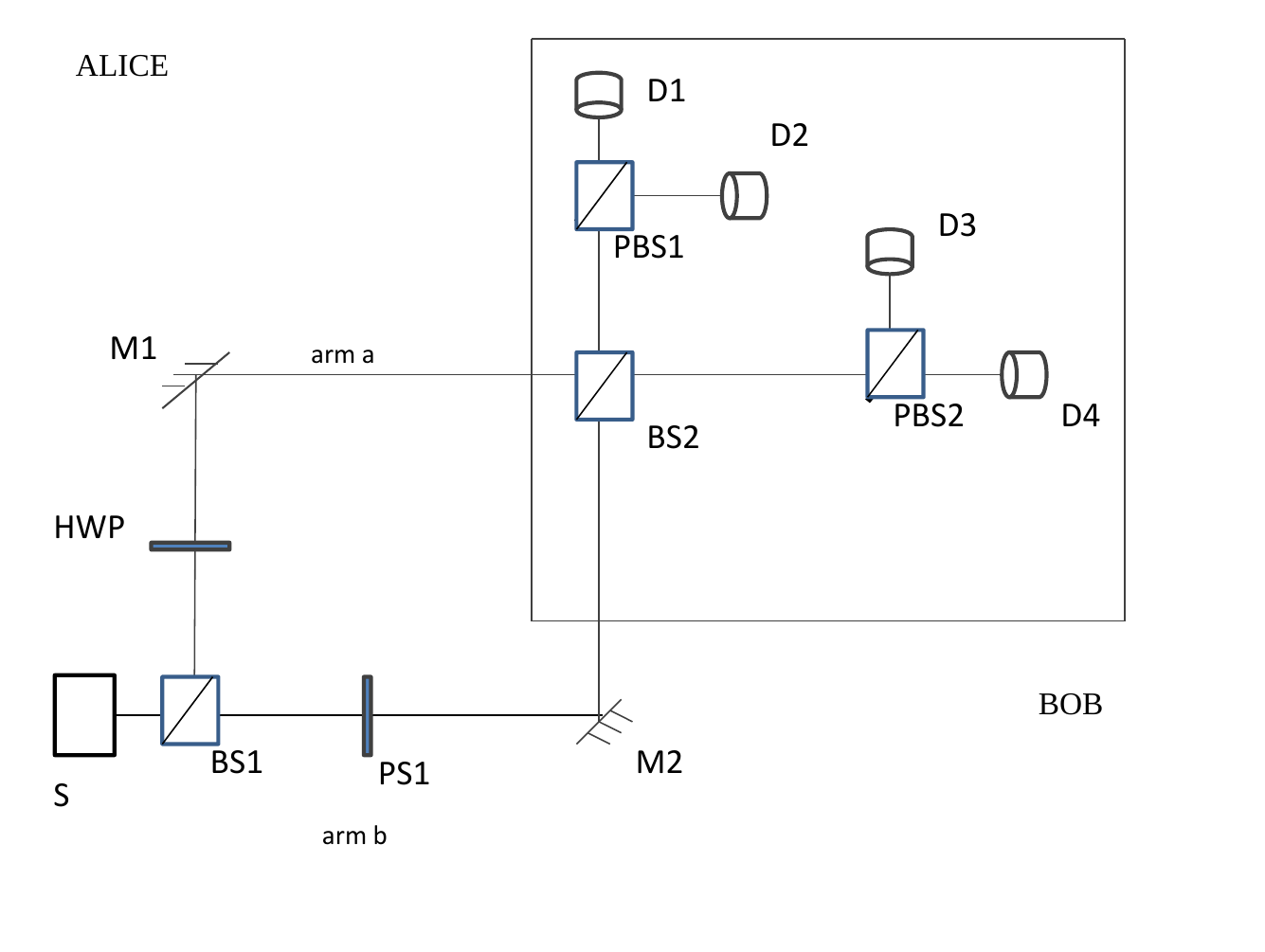}
\caption{BB84 set-up:  Alice transmits  a state to  Bob in one  of the
  bases $G_j^A$  by suitably applying  the linear optical  elements of
  beam splitters, HWP, QWP and PS. Bob may recombine the reflected and
  the  transmitted channels at  BS2.  Finally,  Bob performs  path and
  polarization measurements  using the polarizing  beam splitters PBS1
  and PBS2.}
\label{fig:bb84}
\end{figure}

Let  us  consider a  photon  that  is  initially polarized  along  the
vertical direction  (its state  denoted by $|0\rangle$).   Taking into
consideration   its   path   (or   position)  variables,   the   joint
path-polarization state can be written as
\begin{equation}
|\psi_{0}\rangle_{ps}=|V\rangle_{s} \otimes |\psi_{0}\rangle_{p}
\label{pathspinpro}
\end{equation}
where the subscripts $p$ and $s$ refer to the path and the spin (i.e.,
polarization)   variables  respectively.   A   photon  in   the  state
$|\psi_{0}\rangle_{ps}$  with Alice,  is incident  on a  beam splitter
(BS1),   whose   transmission   and   reflection   probabilities   are
$|\alpha|^{2}$      and     $|\beta|^{2}$      respectively,     where
$|\alpha|^{2}+|\beta|^{2}=1$ (cf. Figure \ref{fig:bb84}).

The  reflected  and transmitted  states  from  BS1  are designated  by
$|\psi_{R}\rangle$  and  $|\psi_{T}\rangle$,  respectively.   Here  we
recall that for any given  lossless beam splitter, arguments using the
unitarity condition show  that for the particles incident  on the beam
splitter, the  phase shift between  the transmitted and  the reflected
states of the particle is $\frac{\pi}{2}$. Note that the beam splitter
acts only on the  path-states without affecting the polarization state
of the particles.

The state of a particle emergent from BS1 can then be written as
\begin{eqnarray}
|\psi_{0}\rangle_{ps} \rightarrow   |\psi_{1}\rangle_{ps}  =  |V\rangle_s
\otimes (\alpha|\psi_{T}\rangle_{p}+i\beta|\psi_{R}\rangle_{p}),
 \label{pathspinpro1}
\end{eqnarray}
where 
\begin{eqnarray}
&&|\psi_{T}\rangle_{p}\equiv \left(%
\begin{array}{c}
  0   \\
  1  \\
  \end{array}%
\right),|\psi_{R}\rangle_{p}\equiv \left(%
\begin{array}{c}
  1   \\
  0  \\
  \end{array}%
\right) {}\nonumber\\&&
|V\rangle_{s}\equiv |0\rangle_{s}\equiv \left(%
\begin{array}{c}
  0   \\
  1  \\
  \end{array}%
\right),|H\rangle_{s}\equiv |1\rangle_{s}\equiv \left(%
\begin{array}{c}
  1   \\
  0  \\
  \end{array}%
\right) \label{notations}
\end{eqnarray}
Our simplest basis, called  $G_1^A$, can be generated without using
the beam splitter:
\begin{eqnarray}
|\Psi_+\rangle &=& |0\rangle_{s}\otimes|\psi_{T}\rangle_{p}
\nonumber\\
  |\Psi_-\rangle &=& |1\rangle_{s}\otimes|\psi_{T}\rangle_{p};
\nonumber\\
|\Psi^\ast_+\rangle &=& |0\rangle_{s} \otimes |\psi_{R}\rangle_{p};
\nonumber\\
  |\Psi^\ast_-\rangle &=& |1\rangle_{s}\otimes|\psi_{R}\rangle_{p}.
\label{fourstate0}
\end{eqnarray}

A basis consisting of  path-polarization entangled elements, and which
is  mutually   unbiased  with  $G_1^A$,  is  $G_2^A$,   given  in  Eq.
(\ref{fourstate}).   It  is  produced   by  a  linear  optical  set-up
consisting of a beam splitter, a half-wave plate (HWP), a quarter-wave
plate (QWP)  and a phase shifter (PS).   For example, $|\Phi_+\rangle$
is produced  from $|\Psi_+\rangle$, by passing the  particle through a
50:50 beam  splitter (BS1, in figure \ref{fig:bb84}),  applying HWP on
the  transmitted wave packet  $|\psi_{T}\rangle_{p}$, followed  by the
application of QWP on both arms. The HWP has the action $|H\rangle_{s}
\leftrightarrow |V\rangle_{s}$.
\begin{eqnarray}
|\Phi_{\pm} \rangle &=& \frac{1}{\sqrt{2}}\left(\frac{|0\rangle_{s}
 \pm |1\rangle_{s}}{\sqrt{2}}
  \otimes|\psi_{T}\rangle_{p} \pm i
  \frac{|0\rangle_{s}-|1\rangle_{s}}{\sqrt{2}}\otimes|\psi_{R}\rangle_{p}\right),\nonumber\\
  |\Phi^\ast_\pm\rangle &=& \frac{1}{\sqrt{2}}\left(\frac{|0\rangle_{s}-|1\rangle_{s}}{\sqrt{2}}
  \otimes|\psi_{T}\rangle_{p} \pm i
  \frac{|0\rangle_{s}+|1\rangle_{s}}{\sqrt{2}}\otimes|\psi_{R}\rangle_{p}\right).
\label{fourstate}
\end{eqnarray}
The bases $G_1^A$ and $G_2^A$  are mutually unbiased in the sense that
any element  in either basis  is an equal weight  superposition (apart
from phase factors) of elements of the other basis.

In dimension $d=4$, there  are $d+1=5$ mutually unbiased bases (MUBs).
Another  mutually  unbiased   entangled  basis,  denoted  $G_3^A$,  in
addition to set (\ref{fourstate}), is:
\begin{eqnarray}
|\Lambda_{\pm} \rangle &=& \frac{1}{\sqrt{2}}\left(\frac{|0\rangle_{s}
 \pm i|1\rangle_{s}}{\sqrt{2}}
  \otimes|\psi_{T}\rangle_{p}+i
  \frac{|0\rangle_{s}-i|1\rangle_{s}}{\sqrt{2}}\otimes|\psi_{R}\rangle_{p}\right), \nonumber\\
  |\Lambda^\ast_\pm\rangle &=& \frac{1}{\sqrt{2}}\left(\frac{|0\rangle_{s}-i|1\rangle_{s}}{\sqrt{2}}
  \otimes|\psi_{T}\rangle_{p} \pm i
  \frac{|0\rangle_{s}+i|1\rangle_{s}}{\sqrt{2}}\otimes|\psi_{R}\rangle_{p}\right).
\label{fourstate1}
\end{eqnarray}
Two others  (separable state) MUBs,  which may be denoted  $G_4^A$ and
$G_5^A$,  can be  produced by  applying  $H \otimes  H$ and  $H^\prime
\otimes H^\prime$  to the elements of  basis (\ref{fourstate0}), where
$H  \equiv \frac{1}{2}(\sigma_z +  \sigma_x)$, while  $H^\prime \equiv
\frac{1}{2}(\sigma_z + \sigma_y)$.

Alice's  states are  analyzed in  Bob's system,  consisting of  a beam
splitter (BS2), followed by  polarization analyzer in each output arm.
For   example,   if   she   sends  the   state   $|\Phi_+\rangle$   or
$|\Phi_{-}\rangle$,  then   after  emerging  from   BS2  (cf.   Figure
\ref{fig:bb84}), the corresponding resulting  states at Bob's site are
given by
\begin{eqnarray}
  |\Phi^\prime_+\rangle&=&\frac{1}{\sqrt{2}}(i|\chi_{1}\rangle\otimes|\psi_{T}'\rangle+
  |\chi_{2}\rangle\otimes|\psi_{R}'\rangle){},\nonumber\\
  |\Phi^\prime_{-}\rangle&=&\frac{1}{\sqrt{2}}(i|\chi_{2}\rangle\otimes|\psi_{T}'\rangle+
  |\chi_{1}\rangle\otimes|\psi_{R}'\rangle),
 \label{beamsplitter2}
\end{eqnarray}
where
$|\chi_{1}\rangle=|0\rangle_{s}$,     $|\chi_{2}\rangle=|1\rangle_{s}$,
$|\psi_T\rangle   =   \frac{1}{\sqrt{2}}\left(|\psi^\prime_T\rangle   +
i|\psi_R^\prime\rangle\right)$       and       $|\psi_T\rangle       =
\frac{1}{\sqrt{2}}\left(|\psi^\prime_T\rangle                          -
i|\psi_R^\prime\rangle\right)$.

\section{Path-polarization entanglement and 
side-channel attacks}

Like  the  BB84  protocol,  our  path-polarization  entangled  version
involves  Alice and  Bob  agreeing on  symbols  0, 1,  2  and 3  being
represented by the elements of each of the five bases mentioned above,
given     by     states     (\ref{fourstate0}),     (\ref{fourstate}),
(\ref{fourstate1}),  etc. We  will consider  two protocols:  one which
employs  only of  the  two MUBs,  to  parallel BB84;  the other  which
includes the full suite of 5 MUBs.
\begin{description}
\item[State   preparation  and   transmission:]   Alice  prepares   an
  intra-particle  path-polarization entangled states  from one  of the
  agreed upon bases, and sends that to Bob.

\item[Bob's  measurement:] Bob uses  mirrors, phase-shifters  and beam
  splitters (Fig.  \ref{fig:bb84})  to measure the transmitted photon,
  choosing  randomly  the   measurement  setting  $G_j^B$  basis,  the
  `primed' versions of  $G_j^A$. He obtains a 2-bit  outcome, which he
  records.
  
\item[Key generation:] The experiment described in the above two steps
  is repeated  many times.  Alice  then declares via  an authenticated
  classical channel the  basis (but not the basis  element) from which
  she chooses the state  randomly.  (The existence of an authenticated
  channel between Alice and Bob, which  gives Bob an edge over Eve, is
  essential to the security of  QKD.) After listening to Alice's group
  declaration,  Bob informs  Alice which  experimental data  should be
  kept and which  ones should be discarded.  The  retained data can be
  potentially used to  distil the secret key, provided  they find that
  the error rate is sufficiently low.

\item[Classical   post-processing.]   Alice   and   Bob  perform   key
  reconciliation  over   the  authenticed  channel,   to  improve  the
  correlation of their respective copy  of the key.  They then perform
  privacy amplification to minimize Eve's information on the key.

\end{description}

The use of  entanglement not only enhances the  bit rate per particle,
but also  protects the protocol against some  side-channel attacks, as
discussed below.  Suppose Eve, the manufacturer of some of the devices
used in  Alice's lab rigs  them in a  way that they  leak \textit{side
  information}  about their  action.   For example,  the tiny  angular
momentum acquired by the QWP through recoil during its rotation of the
photon   polarization,  can   in   principle  be   electromagnetically
transmitted  outside the  lab and  monitored by  Eve  located outside.
This side  channel can reveal not  only Alice's basis  choice but also
the  outcome information,  thereby entirely  compromising  a practical
realization of BB84-like protocols.

The attack can be mathematically represented as follows:
\begin{eqnarray}
|b\rangle|A\rangle_D                     &\rightarrow&
|b\rangle|A\rangle_D                          ~(b=0,1)
 \nonumber\\             |0\rangle|P\rangle_D|0\rangle_\varphi           &\rightarrow&
|{+}\rangle|P_+\rangle_D|0\rangle_\varphi     \rightarrow
|{+}\rangle|P\rangle_D|{+}\rangle_\varphi \nonumber \\ 
|1\rangle|P\rangle_D|0\rangle_\varphi  
 &\rightarrow&           |{-}\rangle|P_-\rangle_D|0\rangle_\varphi            \rightarrow
|{-}\rangle|P\rangle_D|{-}\rangle_\varphi
\label{eq:scha}
\end{eqnarray}
where $|A\rangle,  |P\rangle$ correspond to states  of initial absence
or presence  of some  device $D$, here  the QWP;  $|P_\pm\rangle$, the
recoiled  state of  the device,  carrying  a small  amount of  angular
momentum acquired when the photon in a $V/H$ state is transformed into
one of the circular polarization states $\frac{1}{\sqrt{2}} (|H\rangle
\pm |V\rangle)$;  $|0\rangle_\varphi, 
|{\pm}\rangle_\varphi$ are the  vacuum state and
state of the electromagnetic leaking channel, produced when the device
relaxes back  from $|P_\pm\rangle$  to its initial  state.  It  may be
noted that the distinction  between $|P_+\rangle$ and $|P_-\rangle$ is
imposed not by unitarity but by angular momentum conservation.

Clearly,  the above  attack renders  standard single-qubit  based BB84
fully insecure. For example, in applying the QWP on state $|0\rangle$,
Eve effects the transformation
\begin{equation}
|0\rangle|P\rangle_D|0\rangle_\varphi \rightarrow 
\frac{1}{\sqrt{2}}(|0\rangle + |1\rangle)|P_+\rangle|0\rangle_\varphi
\rightarrow 
\frac{1}{\sqrt{2}}(|0\rangle + |1\rangle)|P\rangle|{+}\rangle_\varphi.
\end{equation}
Eve thus obtains not only  the basis but also outcome information.  On
the other hand, suppose Eve attempts this attack on the above protocol
based on intra-particle entanglement. She leaves the group $G_{1}^{A}$
unaffected,  but $G_{2}^{A}$  is  modified.  For  example, in  Alice's
attempt  to  prepare  the  state  $|\Phi_+\rangle$,  she  effects  the
transformation:
\begin{eqnarray}
\frac{1}{\sqrt{2}}\left(|0\rangle|\psi_T\rangle                       +
i|1\rangle|\psi_R\rangle\right)    |P\rangle_D   |0\rangle_\varphi   &
\rightarrow                                                           &
\frac{1}{\sqrt{2}}\left(\frac{|0\rangle_{s}+|1\rangle_{s}}{\sqrt{2}}|{+}\rangle_\varphi
\otimes|\psi_{T}\rangle_{p}     +      i     \frac{|0\rangle_{s}     -
  |1\rangle_{s}}{\sqrt{2}}|{-}\rangle_\varphi                   \otimes
|\psi_{R}\rangle_{p}\right)|P\rangle_D \nonumber \\
&                              \ne &
|\Phi_+\rangle|P\rangle_D|0\rangle_\varphi
\end{eqnarray}
It   is   easily   shown   that  as   as   $|{+}\rangle_\varphi$   and
$|{-}\rangle_\varphi$ become  more distinguishable, the intra-particle
entanglement  decreases,  which  is  a manifestation  of  monogamy  of
entanglement.       In      particular,     if      $_\varphi{\langle}
{+}|{-}\rangle_\varphi \equiv \cos(\theta)$,  then the concurrence (of
entanglement)  of the state  received by  Bob is  $\cos^2(\theta)$, an
effect which can be detected  by Bob by testing for nonlocality.  This
behavior is a reflection of the monogamy property of nonlocality.


\section{Security check \label{sec:cheq}}

We model  Eve's attack as  a simple intercept-resend attack  on single
particles, where  she measures in  one of the legitimate  bases, which
are  assumed to  be  publicly  known.  Eve  can  also eavesdrop  their
authenticated  classical channel,  and thus  make use  of  their basis
announcements.  Eve's strategy is to measure the particles randomly in
one of the  legitimate bases.  She forwards the  measured state to Bob
in negligible time, and waits until after their public announcement of
bases to find out when she got it right.

Without   loss  of   generality,   suppose  Alice   sends  the   state
$|\Phi_+\rangle$ and Eve attacks  fraction $f$ of particles from Alice
to Bob.   Eve has an equal chance  of measuring in the  right or wrong
basis.   If she  measures  in $G_2^A$  (with  probability $f/2$),  she
always obtains  $|\Phi_+\rangle$, which  she forwards to  Bob, without
introducing any error.

On the other hand, if she  measures in a basis other than $G_2^A$, she
finds any one of the  four basis elements with equal probability.  She
forwards the obtained state to Bob.  After Alice's public announcement
of basis, she is equally unsure of what state Alice prepared as she is
of what state  Bob obtained. The error rate $e$  generated is given by
the probability that Alice and  Bob, measuring in the same basis, find
the wrong  outcome, which is, respectively for  the two-basis protocol
and the five-basis protocol:
\begin{subequations}
\begin{eqnarray}
e &=&  \frac{f_2}{2} \times \frac{3}{4} = \frac{3f_2}{8}
\Longrightarrow f_2 = \frac{8e}{3}  \label{eq:e1} \\
e&=& \frac{4f_5}{5} \times \frac{3}{4} = \frac{3f_5}{5}
 \Longrightarrow f_5 = \frac{5e}{3}. 
\label{eq:e2}
\end{eqnarray}
\label{eq:e}
\end{subequations}

Eve's average information (symmetrically with respect to Alice or Bob)
per transmitted particle is given for the two protocols as:
\begin{subequations}
\begin{eqnarray}
I_2(A:E)   &=&  I_2(B:E)   =   2 \times \left(\frac{f_2}{2}  +   
\frac{f_2}{2}\frac{1}{4}\right)
= \frac{10f_2}{8} = \frac{10e}{3} ~\textrm{bits}. \label{eq:IABa} \\
I_5(A:E)   &=&  I_5(B:E)   =   2 \times \left(\frac{f_5}{5}  +   
\frac{4f_5}{5}\frac{1}{4}\right)
= \frac{4f_5}{5} = \frac{4e}{3} ~\textrm{bits}. \label{eq:IABb}
\end{eqnarray}
\label{eq:IAE}
\end{subequations}
Because  of the mutual  unbiasedness property  between any  two bases,
after Alice and Bob have  reconciled their bases, Eve's action induces
on any  input symbol $m$,  the output probability  distribution $P(n)$
where $P(n=m) = 1-e$ and $P(n\ne m) = \frac{e}{3}$.  The corresponding
Shannon  entropy  functional  is  given by  $H\left(1-e,  \frac{e}{3},
\frac{e}{3}, \frac{e}{3}\right) = -(1-e)\log_2(1-e) -e\log(e/3)$.

Assuming Alice sends all 4 states in all bases with equal probability,
Bob's information is given by the mutual information:
\begin{eqnarray}
I(A:B) &=& H(B) - H(B|A) \nonumber \\
       &=& 2 - H\left(1-e,\frac{e}{3},\frac{e}{3},\frac{e}{3}\right).
\label{eq:IAB}
\end{eqnarray}
The condition for a positive key rate in a protocol is that
\begin{equation}
K_\alpha = I(A:B) - \min\{I_\alpha(A:E), I_\alpha(B:E)\} > 0,
\hspace{0.5cm} \alpha \in \{2,5\},
\label{eq:poskey}
\end{equation}
whose sign  is determined  by Eqs.  (\ref{eq:IAB})  and (\ref{eq:IAE})
\cite{ck78}.   $K_j$ is  a  measure of  the  secret bits  that can  be
distilled after  Alice and Bob perform key  reconciliation and privacy
amplification.

\begin{figure}
\includegraphics[width=14.0cm]{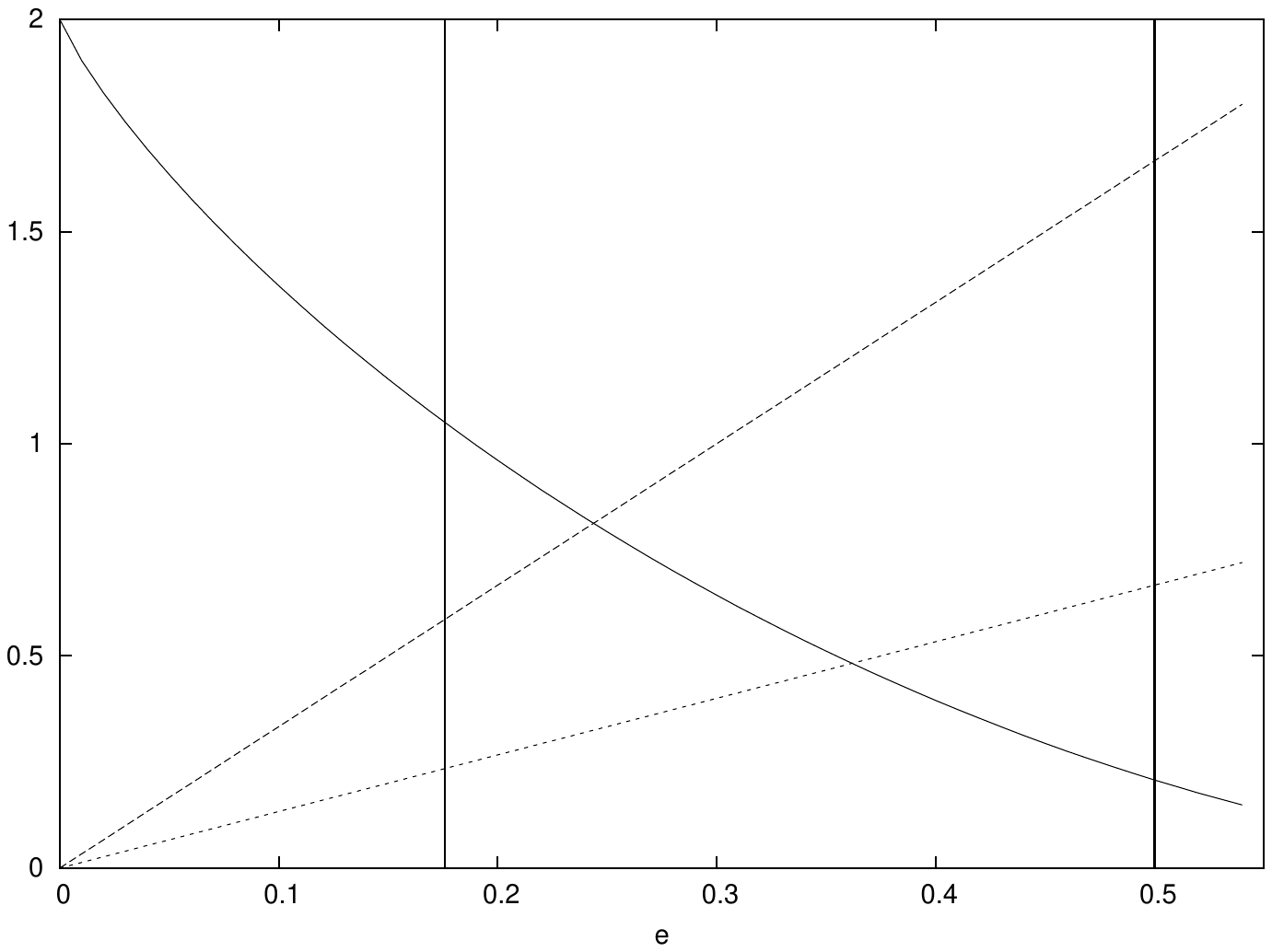}   
\caption{The descending  curve is Bob's  information as a  function of
  error rate  $e$ (Eq.  (\ref{eq:IAB})).  The  upper (lower) ascending
  line  Eve's  information (Eq.  (\ref{eq:IAE}))  when  Alice and  Bob
  choose  two (all  five) mutually  unbiased bases  for  preparing the
  input symbol states.  The tolerable  error rate is about 24\% (36\%)
  when two (five) bases are used for encoding.  For Eve's depolarizing
  action Eq. (\ref{eq:wernere}), the noisy  state is nonlocal for $e <
  e_{\rm  local} \approx 0.17$  (Eq.  (\ref{eq:eloc}),  first vertical
  line)   and  entangled   for  $e   <   e_{\rm  ent}   =  0.5$   (Eq.
  (\ref{eq:eent}), second  vertical line).  Thus  the intercept-resend
  attack on  both protocols allows  secure states that are  local, but
  never secure states that are disentangled.}
\label{fig:ir}
\end{figure}

Bob's  and   Eve's  information  on   Alice  are  plotted   in  Figure
\ref{fig:ir}  both  for  the  two-basis  as  well  as  the  five-basis
protocols.  The  tolerable error rate  for the former  (latter), where
$I_\alpha(A:E)$  just exceeds  $I(A:B)$, is  found to  be  about $24\%
\equiv  e_2$ $(36\%  \equiv e_5)$.   If Eve  launches a  more powerful
attack,  that  is,  one   that  increases  $I_\alpha(A:E)$  for  fixed
$I(A:B)$, then clearly the tolerable  error rate in each protocol will
be lesser.

By Eve's  interference, she is  acting as a depolarizing  channel
that has the action:
\begin{subequations}
\begin{eqnarray}
\rho_j^k         \longrightarrow         \mathcal{E}_2(\rho_j^k)         &=&
\left(1-\frac{f_2}{2}\right)\rho_j^k                                    +
\frac{f_2}{2}\frac{I_4}{4},  \label{eq:wernera} \\
\rho_j^k         \longrightarrow         \mathcal{E}_5(\rho_j^k) 
&=&\left(1-\frac{4f_5}{5}\right)\rho_j^k                                    +
\frac{4f_5}{5}\frac{I_4}{4},
\label{eq:wernerb}
\end{eqnarray}
\label{eq:werner}
\end{subequations}
where  $k$ $(\in  \{1,  2,3,4,5\})$  labels the  basis  and $j$  $(\in
\{1,2,3,4\})$  labels  the  basis  elements.  If  Alice  transmits  an
entangled basis  element, then the  state in Eq.  (\ref{eq:werner}) is
(up to local unitaries) a Werner state \cite{ppt}.

Substituting   for   $f_\alpha$  from   Eq.    (\ref{eq:e1})  in   Eq.
(\ref{eq:wernera}),     or      from     Eq.      (\ref{eq:e2})     in
Eq. (\ref{eq:wernerb}), we find Eve's depolarizing channel in terms of
error rate:
\begin{equation}
\rho_j^k         \longrightarrow        \mathcal{E}_\alpha(\rho_j^k)        =
\left(1-\frac{4e}{3}\right)\rho_j^k                                   +
\frac{4e}{3}\frac{I_4}{4}, \hspace{0.5cm}
\alpha \in \{2,5\}.
\label{eq:wernere}
\end{equation}
If Alice  sends a  separable basis element,  then Eve's  attack, which
depolarizes Alice's  state, does not affect the  entanglement.  On the
other    hand,    suppose    the    input    is    $\rho^2_1    \equiv
|\Phi_+\rangle\langle\Phi_+|$,  so that  Bob receives  (up to  a local
unitary) a Werner state as determined by Eq.  (\ref{eq:werner}).

It  is known  that Bell-type  inequalities for  multipartite separable
states  are  exponentially  stronger than  local-realism  inequalities
\cite{roy04}.  In  the present case,  let the measurement  settings on
the first and second qubit  (i.e., the polarization and path qubit) be
given by the following directions:
\begin{eqnarray}
\vec{a}_{1}=\hat{i}, \vec{a}_{2}=\hat{j},
\vec{a}_{3}=\frac{1}{\sqrt{2}}\hat{i}+\frac{1}{\sqrt{2}}\hat{j},\nonumber\\
\vec{b}_{1}=\frac{1}{\sqrt{2}}\hat{i}+\frac{1}{\sqrt{2}}\hat{j},
\vec{b}_{2}=\frac{-1}{\sqrt{2}}\hat{i}+\frac{1}{\sqrt{2}}\hat{j},
\vec{b}_{3}=\hat{j}
\label{eq:settings}
\end{eqnarray}
which are used for evaluating one of the following Bell correlations
\begin{equation}
S  =  E(\vec{a}_{1},  \vec{b}_{1})  +  E(\vec{a}_{2},  \vec{b}_{1})  +
E(\vec{a}_{1}, \vec{b}_{2}) -E(\vec{a}_{2},\vec{b}_{2}).
\label{eq:bellop}
\end{equation}
It is readily shown that for local-realist models, $S \le S_{LR} = 2$.
The correlation for the singlet is given by
\begin{eqnarray}
E(\vec{a}_{i},\vec{b}_{j})=-(\vec{a}_{i},\vec{b}_{j}),
\label{eq:corr}
\end{eqnarray}
so that $S = -2\sqrt{2} = \sqrt{2}S_{LR}$.
The most general separable state is given by
\begin{eqnarray}
\rho_{sep}=\int\int\sigma(\vec{n}_{a},\vec{n}_{b})|n_{a}\rangle\langle
n_{a}|\otimes |n_{b}\rangle\langle n_{b}|d\vec{n}_{a}d\vec{n}_{b},
\label{eq:rhosep}
\end{eqnarray}
where
$\int\int\sigma(\vec{n}_{a},\vec{n}_{b})d\vec{n}_{a}d\vec{n}_{b}=1$
and
\begin{eqnarray}
\vec{n}_{a}=\sin\theta_{a}\cos\phi_{a}\hat{i}                         +
\sin\theta_{a}\sin\phi_{a}\hat{j}+
\cos\theta_{a}\hat{k}{}\nonumber\\            \vec{n}_{b}            =
\sin\theta_{b}\cos\phi_{b}\hat{i}+\sin\theta_{b}\sin\phi_{b}\hat{j}+
\cos\theta_{b}\hat{k}
\end{eqnarray}
The correlations for $\rho_{sep}$ can be calculated as
\begin{eqnarray}
E(\vec{a}_{i},\vec{b}_{j})= \textrm{Tr}[\rho_{sep}
\vec{\sigma}.\vec{a}_{i}\otimes\vec{\sigma}.\vec{b}_{j}]
\end{eqnarray}
Using   equations    Eqs.    (\ref{eq:settings}),   (\ref{eq:bellop}),
(\ref{eq:corr}) and  (\ref{eq:rhosep}), the  upper and lower  bound of
the quantity $S$ in Eq. (\ref{eq:bellop}) is given by
\begin{eqnarray}
S &=&\sqrt{2}\int\int\int\int\sigma(\theta_{a},\theta_{b},\phi_{a},\phi_{b})
\sin^{2}\theta_{a}\sin^{2}\theta_{b}\sin(\phi_{a}+\phi_{b})
d\theta_{a}d\theta_{b}d\phi_{a}d\phi_{b}{}\nonumber\\
&\Rightarrow& -\sqrt{2}\leq S \leq \sqrt{2},
\label{eq:bellbound}
\end{eqnarray}
so  that  quantum  bound  for  separable states,  $S_{\rm  max:sep}  =
\sqrt{2} <  S_{LR}$.

Each element  of the entangled bases  is locally equivalent  to a Bell
state,  and yields  a  Bell inequality  violation  of $2\sqrt{2}$  for
suitable  settings,  whereas  $S(I_4)  =  0$.   After  Alice's  public
announcement of bases, she and  Bob divide the transmitted states into
sub-ensembles corresponding to each symbol state.

In  view   of  Eq.   (\ref{eq:wernere}),  for  a   given  sub-ensemble
corresponding to an entangled symbol state, Bob observes:
\begin{equation}
S = \left(1 - \frac{4e}{3}\right)2\sqrt{2}.
\label{eq:vio}
\end{equation}
This is nonlocal when $\langle S\rangle > S_{LR} = 2$, or
\begin{equation}
e  <  \frac{3}{4}\left(1  -  \frac{1}{\sqrt{2}}\right)  \equiv  e_{LR}.
\label{eq:eloc}
\end{equation}
From  Eq. (\ref{eq:wernere}),  setting  $\left(1-\frac{4e}{3}\right) >
\frac{1}{3}$   as   the  necessary   and   sufficient  condition   for
entanglement  \cite{werner} of  Werner states  by  the Peres-Horodecki
positive-partial-transpose  criterion \cite{ppt},  we find  that Bob's
states are entangled when
\begin{equation}
e < \frac{1}{2} \equiv e_{\rm ent}.
\label{eq:eent}
\end{equation}
Since $e_2  > e_{LR}$ and  $e_5 > e_{LR}$,  it follows that  there are
local states  that allow secrecy  extraction for both  protocols under
the considered attack.

The  corresponding values  of $S$  are, from  Eq.  (\ref{eq:wernere}),
$S_2  = \left(1  -  \frac{4e_2}{3}\right)2\sqrt{2} \approx  1.36S_{\rm
  max:sep}$, and similarly $S_5 \approx 1.04S_{\rm max:sep}$, implying
that  $e_2  <  e_{\rm  ent}$  and  $e_5 <  e_{\rm  ent}$  (cf.  Figure
\ref{fig:ir}).  In other words, all  states under the protocol for the
given  class   of  attacks  are  necessarily   entangled.   Since  the
considered  attacks are  clearly not  the strongest  possible  for the
protocols  considered, this  implies  that security  or  secrecy is  a
strictly stronger  condition than entanglement (in  that more powerful
attacks will  reduce the tolerable  error rate, and thus  increase the
amount entanglement  in the state  at the security threshold).   It is
interesting to note that in the case of the BB84 protocol, interpreted
as an Ekert91-like protocol,  the concepts of nonlocality and security
coincide  \cite{scagis1,scagis2},  whereas  in the  device-independent
scenario,   security  is   a  stronger   condition   than  nonlocality
\cite{acin07}.

The  above  analysis   cryptographically  illustrates  the  difference
between  nonlocality and  entanglement. In  particular, it  shows that
security,     nonlocality     and     entanglement    are     distinct
resources. Security is more  aligned with privacy of randomness, which
is known to be distinct from nonlocality and entanglement \cite{amp}.

\section{Conclusions}

We have proposed the use of intra-particle entanglement in a BB84-like
protocol for realizing QKD.  This has the virtue of enhancing security
against  a  class  of   side-channel  attacks,  while  being  easy  to
experimentally  implement,  as it  requires  only  linear optics.   In
particular, an attack that reveals side-information about the photon's
angular momentum  would be undetected in a  standard implementation of
BB84, whereas our intra-particle entanglement reveals the attack via a
reduction in correlation.

The security of two protocols,  one using two mutually unbiased bases,
and  another  using  five,  are  analyzed under  a  simple  incoherent
attack. We  find that  secrecy is a  strictly stronger  condition than
entanglement.  On  the other  hand, there are  some local  states that
allow   extraction   of   secret   bits.   Our   results   demonstrate
cryptographically  that  nonlocality,  entanglement  and  secrecy  are
related but inequivalent concepts, the last being more akin to privacy
of randomness.

\section*{Acknowledgement}

ASM and DH acknowledge  support from the DST Project SR/S2/PU-16/2007.
DH also thanks the Centre for Science, Kolkata for support.

\end{document}